# Preparation and Characterization of Nano-particle Substituted Barium Hexaferrite

Yomen Atassi[*], Iyad Seyd Darwish and Mohammad Tally

Department of applied physics, Higher Institute for Applied Sciences and Technology, HIAST, P.O. Box 31983, Damascus, Syria.

**Abstract**: High density magnetic recording requires high coercivity magnetic media and small particle size. Barium hexaferrite $BaFe_{12}O_{19}$ has been considered as a leading candidate material because of its chemical stability, fairly large crystal anisotropy and suitable magnetic characteristics. In this work, we present the preparation of the hexagonal ferrite $BaFe_{12}O_{19}$ and one of its derivative; the Zn-Sn substituted hexaferrite by the chemical co-precipitation method. The main advantage of this method on the conventional glass-ceramic one, resides in providing a small enough particle size for magnetic recording. We demonstrate using the X-ray diffraction patterns that the particle size decreases when substituting the hexaferrite by the Zn-Sn combination. This may improve the magnetic properties of the hexaferrite as a medium for HD magnetic recording

1. Introduction

Nanosized ferrite materials have attracted great attention in recent years. They exhibit unusual physical and chemical properties significantly different from those of the bulk materials due to their extremely small size and large specific surface area [1,2]. More precisely, some magnetic properties, such as saturation magnetization and coercitivity, depend strongly on the particle size, morphology and microstructure of the materials.

---

[*] corresponding author: yomen.atassi@hiast.edu.sy

On the other hand, High-coercivity magnetic materials with nanoparticle size are required for high density magnetic recording such as high density magnetic tapes and floppy disks. The hexagonal ferrite $BaFe_{12}O_{19}$ constitutes a promising candidate material due to its chemical stability, fairly large crystal anisotropy and suitable magnetic properties. In fact, it has been demonstrated that barium ferrite tapes, which utilize very small particles D=40-50 nm and of high coercivity $H_c = 1750 - 2060\,Oe$ offer superior high-density recording performance [3].

In this work, we present the preparation of the hexagonal ferrite $BaFe_{12}O_{19}$ and one of its derivative; the Zn-Sn substituted hexaferrite by the chemical co-precipitation method. The main advantage of this method on the conventional glass-ceramic one, resides in providing a small enough particle size for magnetic recording. Although, there are different other techniques that may lead to nanosized particles like ball milling, sol-gel and hydrothermal techniques, the coprecipitation technique seems to be the most convenient for the synthesis of nanoparticles because of its simplicity and better control over crystallite size. In the coprecipitation technique, common hydroxides of barium and iron are precipitated from their salts by adding the metal salts solution to an alkaline medium, and crystallized to barium hexaferrite upon suitable heat treatment. The most common drawback of the coprecipitation technique is formation of agglomerated precipitate. The crystalline powders formed from these precipitates are nanocrystalline but agglomerated, which are difficult to disperse.

We report here that using isopropyl alcohol to wash the precipitate may overcome this problem. On the other hand, we demonstrate, using the X-ray diffraction patterns that the particle size decreases when substituting the hexaferrite by Zn-Sn combination. This may enhance the properties of the hexaferrite to meet the requirements of high density magnetic recording.

## 2. Experimental Section

All the used reagents were of analytical grade.

### 2.1. Preparation of barium hexaferrite

Aqueous solutions of barium chloride and iron chloride in appropriate volumetric amounts were used as starting materials in the synthesis of phase pure $BaFe_{12}O_{19}$. Precipitation of the desired powder was then achieved by the technique of acid-base titration.

It was noticed that although a Ba/Fe molar ratio of 1/12 in aqueous solution should be sufficient for the preparation of pure barium ferrite according to stoichiometry, an excess of barium chloride was necessary because barium hydroxide is slightly soluble in water (solubility product at $25°C$, $K_s(Ba(OH)_2.8H_2O) = 2.55 \times 10^{-4}$ [4]). The molar ratio of Ba/Fe in aqueous solution was taken as 1/11. The determination of the appropriate molar ratio was illustrated elsewhere [2,3].

The product of the co-precipitation was filtered, washed repeatedly with deionized water to remove the unwanted chloride ions. The complete removal of chloride ions was confirmed from the pH and conductivity measurements of the deionized water and the wash effluent. After washing, the precipitate was dewatered with isopropyl alcohol to remove the surface adsorbed water molecules, which might lead to the agglomeration of the particles during the drying and further processing. The powder was then dried at a temperature of $110°C$ for 12 hours, and then divided into two patches. The first patch was pressed into pellets of 10 mm diameter and a thickness of 4 mm for dilatometer studies and the second patch was pressed into pellets (of 25 mm at 200 MPa) and toroids (with internal and external diameters 10 and 25 mm, respectively, and a thickness of about 8 mm) and sintered at $800°C$ for two hours.

### 2.2. Preparation of Zn-Sn substituted hexaferrite

Aqueous solutions of barium chloride, iron chloride, zinc chloride and tin chloride (IV) in appropriate volumetric amounts were used as starting materials in the synthesis of phase pure $BaFe_{11.4}Zn_{0.3}Sn_{0.3}O_{19}$. One should notice that the standard redox potentials of iron and tin ions are $E°(Fe^{3+}/Fe^{2+}) = 0.770 eV$ and $E°(Sn^{4+}/Sn^{2+}) = 0.159 eV$. So we use directly tin IV: $Sn^{+4}$ instead of $Sn^{+2}$ in order to avoid the reduction of ferric ions to ferrous ones by $Sn^{+2}$ and the formation of $Sn^{+4}$. The procedure of the preparation of Zn-Sn substituted hexaferrite is identical to the one described above paragraph 2.2.

### 2.3. Densification

In order to determine the suitable sintering temperature, dilatometer studies were performed using (Setaram, TMA 92) at a heating rate of $10°C/min$ in nitrogen.

### 2.4. X-ray diffraction pattern

A computer interface X-ray powder diffractometer (Philips) with Cu K$\alpha$ radiation ($\lambda = 0.1542 nm$) was used to identify the crystalline phase.

The data collection was over the 2-theta range of $10°$ to $100°$ in steps of $0.02°/sec$. The average crystallite size was related to the pure X-ray line broadening using Scherrer's formula [5]. The Fourier method was applied to the (110) and (107) lines.

The mean hexagonal diameter $D$ and thickness $t$ of particles have been determined from the apparent size $\varepsilon_F(110)$ and $\varepsilon_F(107)$ [6]. In fact, $\varepsilon_F$ value is the ratio of the sample volume to the projection area of this volume on the reflecting plane (hkl). This definition allows to deduce the mean hexagonal diameter $D$ from $\varepsilon_F(110)$ [6]:

$$D = \frac{2}{\sqrt{3}} \varepsilon_F(110)$$

The thickness was determined by the relation [6]:

$$t = \varepsilon_F(107) \cos\varphi$$

Where $\varphi$ is the angle between the (001) direction and the perpendicular to the (107) plane, in our case $\varphi = 33°$.

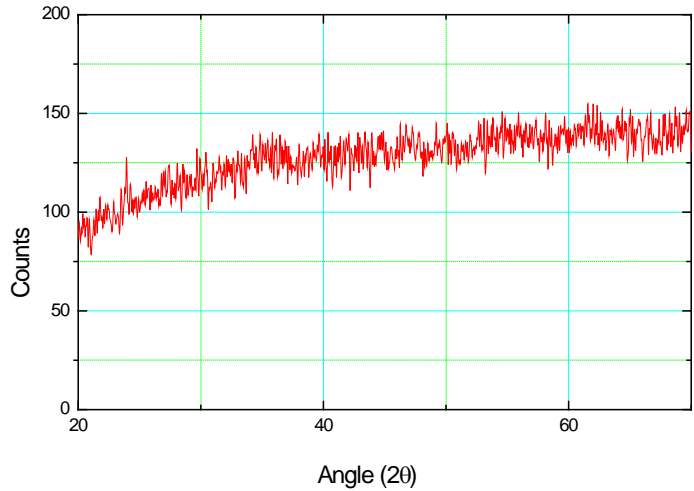

Fig.1: X-ray pattern of barium hexagonal ferrite before heat treatment.

## 3. Results and discussions

### 3.1. Barium hexagonal ferrite

The XRD pattern of the precipitate before heat treatment shows that it's amorphous, Fig. 1.

The formation of barium hexaferrite was confirmed by the XRD pattern after heat treatment and no other phases were apparently detectable, Fig.

2. All peaks matched well with the characteristics reflections of barium hexagonal ferrite, as it's indicated clearly in table 1. The XRD pattern presents a noticeable line broadening, indicating the fine-particle nature of the hexagonal ferrite so prepared. The mean hexagonal diameter and thickness of particles were $D = 45.2 \pm 3.8\,\text{nm}$ and $t = 31.1 \pm 1.5\,\text{nm}$.

| $2\theta_{exp}$ | $2\theta_{th}$ | $h\,k\,l$ | (%) $I_{exp}$ | $I_{th}$ (%) | $d\,(\text{Å})$ |
|---|---|---|---|---|---|
| 32.2 | 32.2 | 107 | 78.5 | 100 | 2.782 |
| 34.1 | 34.1 | 114 | 100 | 98 | 2.627 |
| 37.1 | 37.1 | 203 | 53.2 | 60 | 2.424 |
| 30.3 | 30.3 | 110 | 45.6 | 55 | 2.947 |
| 30.8 | 30.8 | 008 | 12.7 | 32 | 2.900 |

Table 1: The five strongest diffraction peaks of barium hexagonal ferrite.

On the other hand, the elemental analysis of hexaferrite after heat treatment confirms Ba/Fe molar ratio of 1/12.

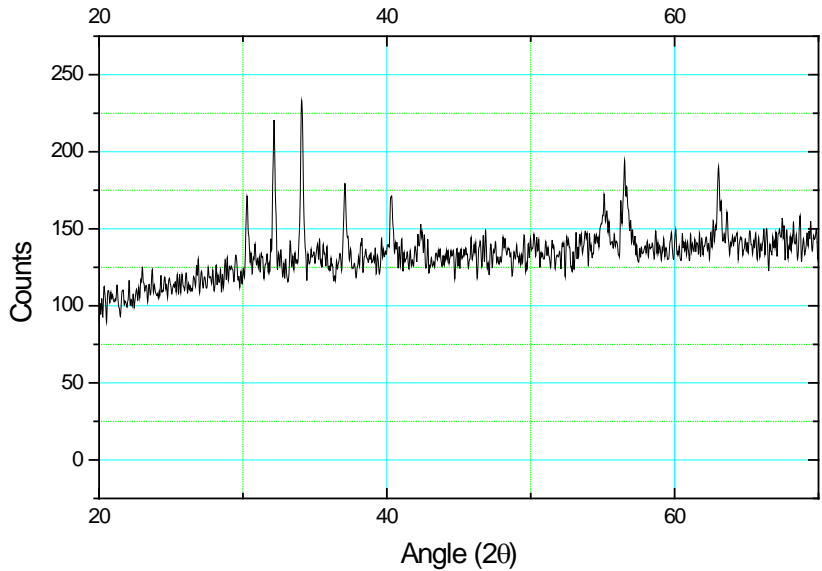

Fig.2: X-ray pattern of barium hexagonal ferrite after heat treatment.

The suitable temperature of sintering was determined using a dilatometer curve, Fig.3.

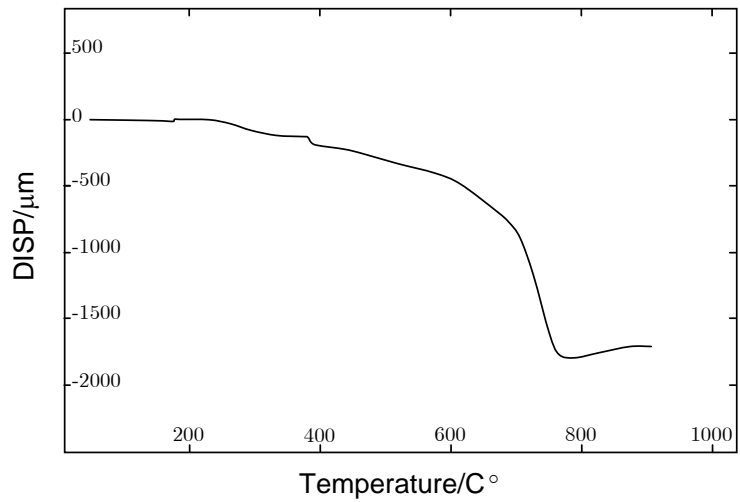

Fig.3: A dilatometer curve of a pellet of the green body.

## 3.2. Zn-Sn substituted barium hexaferrite

The XRD pattern of the precipitate before heat treatment shows that it's also amorphous, Fig. 4.

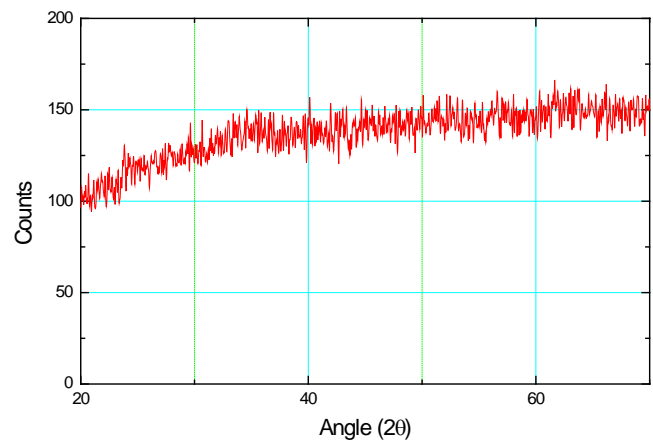

Fig.4: X-ray pattern of Zn-Sn substituted barium hexaferrite before heat treatment.

The formation of the Zn-Sn barium hexaferrite was confirmed by the XRD pattern after heat treatment, Fig. **5**.

All peaks matched well with the characteristics reflections of barium hexagonal ferrite, as it's indicated clearly in table 2. There is still a small residual amount of ferric oxide at $33°$.

| $hkl$ | $I_{exp}$ (%) | $I_{th}$ (%) | $2\theta_{exp}$ | $2\theta_{ref}$ |
|---|---|---|---|---|
| 114 | 100 | 100 | 34 | 34.2 |
| 107 | 84 | 79.3 | 32.1 | 32.1 |
| 2011 | 48 | 69 | 56.2 | 56.1 |
| 217 | 52 | 62.1 | 55 | 54.5 |
| 220 | 52 | 48.3 | 63 | 62.4 |

Table 2: The five strongest diffraction peaks of substituted barium hexferrite.

Substituting hexagonal ferrite by the combination Zn-Sn results in peaks broadening and intensities decreasing. This indicates a decrease in particle size. The mean hexagonal diameter and thickness of particles were $D = 31.0 \pm 1.7$ nm and $t = 15 \pm 0.7$ nm. We notice a decrease in crystallite size when doping the hexaferrite with Zn-Sn, and this is suitable for high density magnetic recording.

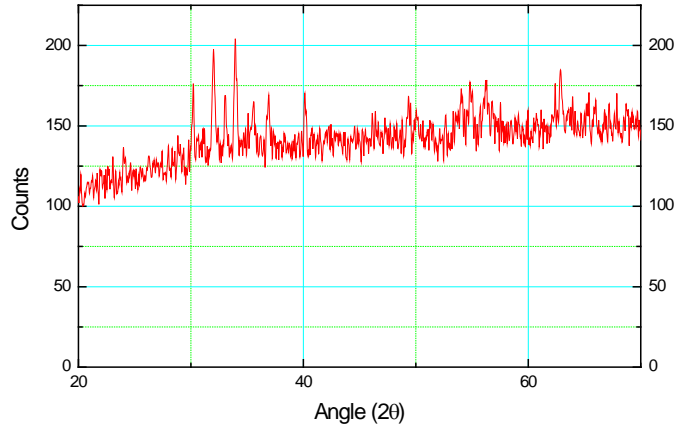

Fig.**5**: X-ray pattern of Zn-Sn substituted barium hexaferrite after heat treatment.

On the other hand, as the radii of the $Zn^{2+}$ ($0.74\overset{\circ}{A}$) and $Sn^{4+}$ ($0.71\overset{\circ}{A}$) ions are larger than that of $Fe^{3+}$ ($0.64\overset{\circ}{A}$), then the lattice parameters a and c are expected to increase when substituting the hexagonal ferrite by the Zn-Sn combination.

This result is confirmed experimentally, as it's clearly indicated in table 3.

|  | $a(\overset{\circ}{A})$ | $c(\overset{\circ}{A})$ | $c/a$ |
|---|---|---|---|
| Hexaferrite | 5.8994 | 23.194 | 3.93 |
| Zn-Sn hexaferrite | 5.9571 | 23.240 | 3.90 |

Table 3: Lattice parameters of barium hexaferrite and its Zn-Sn substitution.

## 4. Summary

Barium hexaferrite and its Zn-Sn derivative have been successfully synthesized by chemical coprecipitation method. Crystallite sizes, measured from X-ray patterns, indicate that it's more suitable for high density magnetic recording applications to use the substituted hexaferrite.

**Acknowledgement:**

The authors would like to thank Dr. Ahmad Alfarra for the fruitful discussions.